\documentclass[aps,prd,onecolumn,groupedaddress]{revtex4}
\usepackage{graphicx}
\usepackage{bm}
\usepackage{epsfig}
\usepackage{color}
\usepackage{colordvi}
\usepackage{amsmath}
\usepackage{epsfig}
\usepackage{mathrsfs}

\newcommand{\beq}{\begin{equation}}
\newcommand{\eeq}{\end{equation}}
\newcommand{\bea}{\begin{eqnarray}}
\newcommand{\eea}{\end{eqnarray}}

\begin{document}

\title{Effective field theory calculation of second post-Newtonian binary dynamics}

\author{James B. Gilmore}\email{james.gilmore@yale.edu}
\affiliation{Department of Physics, Yale University, New Haven,
Connecticut 06520, USA}

\author{Andreas Ross}\email{andreas.ross@yale.edu}
\affiliation{Department of Physics, Yale University, New Haven,
Connecticut 06520, USA}

\date{\today}

\begin{abstract}
We use the effective field theory for gravitational bound states,
proposed by Goldberger and Rothstein, to compute the interaction
Lagrangian of a binary system at the second post-Newtonian order.
Throughout the calculation, we use a metric parametrization based on
a temporal Kaluza-Klein decomposition and test the claim by Kol and
Smolkin that this parametrization provides important calculational
advantages. We demonstrate how to use the effective field theory
method efficiently in precision calculations, and we reproduce known
results for the second post-Newtonian order equations of motion in
harmonic gauge in a straightforward manner.

\end{abstract}

\maketitle

\section{Introduction}

In the last two decades, significant progress has been made towards
the detection of gravitational waves (GWs) via laser interferometry.
Currently, the ground-based experiments LIGO
\cite{Abramovici:1992ah}, VIRGO \cite{Giazotto:1988gw}, GEO
\cite{Willke:2002bs}, and TAMA \cite{Ando:2001ej} are actively
searching for GWs \cite{GWsearch}. Moreover, the proposed LISA
experiment \cite{Danzmann:2003tv}, due to be the first space-based
GW detector, will search for GWs in a complementary frequency band
to the ground-based experiments and is expected to achieve high
event rates at an unprecedented signal-to-noise ratio
\cite{Gair:2004iv}.

A particularly interesting source of GWs, which is expected to be
detected, is the compact binary system undergoing coalescence, with
neutron star (NS) and/or black hole (BH) constituents. Current
experiments have yet to detect the binary inspiral signal. However,
Advanced LIGO \cite{AdvLIGO}, an upgrade of LIGO scheduled to come
online in 2014, may allow for routine detection of such events. This
is due to a $\sim10$-fold increase in sensitivity over LIGO, which
will in turn result in an increase of the accessible event rate by a
factor $\sim1000$. Current estimates for the number of expected
NS/NS, BH/BH, and BH/NS events in Advanced LIGO are roughly
$10-100$, $1-500$, and $1-30$ per year, respectively
\cite{Belczynski:2006zi, Maggiore:1900zz}.

All three stages of the binary coalescence, inspiral, merger, and
ringdown, are potentially detectable. The inspiral phase, where the
characteristic orbital velocity is $v^2 \ll 1$ (in units where
$c=1$), can be computed analytically using an expansion in $v^2 \sim
Gm/r$. The merger is computed numerically \cite{Pretorius:2007nq},
and there has been significant recent progress in this area
\cite{Numerics}. The ringdown can be treated analytically using
quasinormal modes \cite{Kokkotas:1999bd}.

The perturbative calculation of the inspiral phase has been
performed with a variety of methods
\cite{Blanchet:2002av,Futamase:2007zz}. Because of the phase
evolution of the inspiral signal and the ability to measure the
total orbital phase to $\sim 10^{-3}$ over the LIGO bandwidth
\cite{Cutler:1992tc}, these perturbation expansions must be
calculated to high order. If we consider a circular orbit in the
adiabatic approximation, the signal phase $\Phi(\omega)$ is related
to the orbital energy $E(\omega)$ and the radiated power $P(\omega)$
through the relation $d^2\Phi / d \omega^2 \sim (dE/d\omega)/P$. An
accuracy of $\sim 10^{-3}$ in the cumulative orbital phase, over the
LIGO bandwidth, can be achieved if the perturbation expansion is
calculated to $\mathcal O(v^6)$ beyond Newtonian dynamics i.e., at
third post-Newtonian order (3PN) \cite{Maggiore:1900zz,
NeedForHighPN}. This implies that we need to know $E(\omega)$ and
$P(\omega)$ to at least 3PN. Since the conservative dynamics,
described in our formalism by a Lagrangian, gives $E(\omega)$, it
also must be known to 3PN. There is also a need to compute the
perturbative expansion to high order, to allow numerical studies of
the binary inspiral to be compared with the analytic PN expansion
\cite{Boyle:2007ft}.

Recently, Goldberger and Rothstein introduced an effective field
theory (EFT) for nonrelativistic gravitational systems, known as
NRGR \cite{Goldberger:2004jt,Goldberger:2006bd} (see
\cite{Goldberger:2007hy} for a pedagogical introduction). EFTs are
particularly well suited to problems with multiple physical scales,
and the binary inspiral which we consider here, is one such problem.
The hierarchy during inspiral takes the form $r_c\ll r\ll\lambda$,
where $r_c$ is the radius of the compact objects, $r$ is the orbital
separation, and $\lambda$ is the wavelength of the emitted GWs. One
can use this hierarchy to set up a tower of EFTs which
systematically account for effects at each scale. This approach
disentangles the physical effects of different scales, resulting in
more tractable calculations. With the definite power-counting scheme
established by Goldberger and Rothstein \cite{Goldberger:2004jt},
the EFT treatment yields a completely systematic description of the
binary inspiral problem. In particular, divergences which arise by
using point particle sources to represent the compact objects are
well understood in a field theory setting, and the effects due to
the spatial extent of the compact objects can be systematically
parameterized by subleading operators.

State-of-the-art calculations using traditional methods have yielded
the conservative equations of motion and the orbital energy of a
binary with spinless constituents to 3PN order
\cite{2PN,3PNScha,3PNBla,3PNItoh}. The 2PN dynamics of a three-body
system have been obtained in \cite{2pn3body}, but the four-body
dynamics at 2PN are not known in closed form, see, for example,
Appendix D of \cite{Mitchell:2007ea}. Furthermore, the power emitted
by a binary with spinless objects through GWs is known at the 3.5PN
level \cite{rad35} or $\mathcal O(v^7)$ beyond the leading
quadrupole formula. While the EFT approach lags behind the
traditional post-Newtonian methods in terms of high precision
calculations of GW observables for binaries without spin, it has led
to a number of interesting results. Spin was incorporated into the
EFT framework and the next-to-leading order (NLO) spin-spin
potential, which enters at 3PN, has been calculated for the first
time \cite{RafIra:Spin}. Dissipative effects such as absorption by
BH horizons have also been considered \cite{WaIrRa:Absorp}. In
\cite{Chu:2006ce}, the thermodynamics of compactified black holes
were studied, and the Einstein-Infeld-Hoffman (EIH) Lagrangian and
the quadrupole formula were derived in arbitrary dimensions in
\cite{Cardoso:2008gn}. The EFT formalism was extended to the case of
extreme mass-ratio binaries where the leading order (LO) self-force
equation was derived \cite{Chad}. Recently, it has been used in
calculations which include interactions beyond Einstein's general
relativity \cite{Cannella:2008nr, Sanctuary:2008jv}.

Within the EFT method, Kol and Smolkin (KS) suggested that a
temporal Kaluza-Klein parametrization of the metric
\cite{Kol:2007rx} improved the calculational efficiency of NRGR. The
KS parametrization has been shown to reduce the complexity required
to calculate the EIH Lagrangian \cite{Kol:2007bc}. Similar
simplifications were observed in the computation of thermodynamic
properties of compactified black holes \cite{Kol:2007rx}, and in the
calculation of the NLO order spin-spin potential
\cite{Levi:2008nh}.

In this paper, we report the NRGR calculation of the 2PN interaction
Lagrangian for a binary system with nonspinning compact objects.
Existing EFT calculations of binary observables have been obtained
at NLO for potential interactions and at LO for the radiated power.
This work presents the first next-to-next-to-leading order
computation of a GW observable with the EFT method. Since the
complexity of our calculation is more involved, we clearly want to
find an optimum method to perform high precision calculations within
NRGR. Previously, it was not clear how useful the KS parametrization
would be as a computational tool beyond NLO. We address this issue
in detail, and we show how the KS parametrization simplifies our 2PN
calculations. The methods of EFT are used to systematically
determine all relevant Feynman diagrams which contribute at 2PN. In
their evaluation, we encounter Feynman integrals corresponding to
one-loop and two-loop integrals, which are computed using standard
techniques. Our work demonstrates how the EFT method to can be used
to efficiently compute conservative dynamics at high precision in
the PN expansion.

\section{Setup}\label{sec:setup}

Here, we outline the ingredients of our calculation of the 2PN
potential and discuss the simplifications we will employ. In this
section, we do not repeat in entirety the setup of the EFT
description of the binary inspiral problem, but refer the reader to
\cite{Goldberger:2004jt}, \cite{Goldberger:2006bd}, and
\cite{Goldberger:2007hy}. Since we deal with conservative dynamics,
we only need potential modes, and can simply set the radiation modes
to zero.

\subsection{Two-body action}

The purely gravitational action is the usual Einstein-Hilbert action
\begin{equation}\label{eqn:EHact}
 S_{EH} = - 2 m_{Pl}^2 \int d^4x \sqrt{-g} R,
\end{equation}
where our conventions are $m_{Pl}^2 = 1/32 \pi G$, $\eta_{\mu \nu} =
\text{diag}\left[1,-1,-1,-1\right]$, ${R^\mu}_{\nu \alpha \beta} =
\partial_\alpha \Gamma^{\mu}_{\nu \beta} - \partial_\beta
\Gamma^\mu_{\nu \alpha} + \Gamma^\mu_{\lambda \alpha}
\Gamma^\lambda_{\nu \beta} - \Gamma^\mu_{\lambda \beta}
\Gamma^\lambda_{\nu \alpha}$, and $R_{\mu \nu} = {R^\alpha}_{\mu
\alpha \nu}$. In addition, we must also fix the gauge. Our choice is
harmonic gauge
\begin{equation}\label{eqn:gf}
 S_{GF} = m_{Pl}^2 \int d^4x \sqrt{-g} \, \Gamma^\mu \Gamma^\nu g_{\mu
 \nu},
\end{equation}
where $\Gamma^\mu = \Gamma^\mu_{\alpha \beta} g^{\alpha \beta}$. It
differs from the linearized harmonic gauge condition used in
\cite{Goldberger:2004jt} and yields different expressions at the 2PN
level. The advantage of using harmonic gauge is that we can compare
intermediate gauge-dependent results, such as the equations of
motion, with the existing literature \cite{Blanchet:2002av}.

For the gravitational coupling to two massive compact objects,
the worldline action for our binary system is given by
\begin{equation}\label{eqn:pp}
 S_{pp}=-\sum_{N=1}^{2} m_N\int d\tau_N + \cdots=-\sum_{N=1}^{2} m_N\int dt \sqrt{g_{\mu
\nu}(t,\mathbf x_N) \frac{dx_N^\mu}{dt} \frac{dx_N^\nu}{dt}} +
\cdots.
\end{equation}
Here, proper time is given by $d\tau^2 = g_{\mu \nu} dx^\mu dx^\nu$
and it is convenient to use coordinate time $t$ to parameterize the
worldline. The dots denote subleading operators encoding finite size
effects. We will ignore finitesize effects since they first enter at
5PN \cite{Goldberger:2004jt}.

\subsection{Kol-Smolkin variables}

The expansion of the metric around flat Minkowski space in the weak
field limit is commonly parameterized in a Lorentz covariant form
such as
\begin{equation} \label{eqn:hmunumetric}
g_{\mu\nu}=\eta_{\mu\nu}+h_{\mu\nu}/m_{Pl},
\end{equation}
where $h_{\mu\nu}$ is the excitation around Minkowski space. The
expansion of Eq. (\ref{eqn:hmunumetric}) was used in the original
setup by Goldberger and Rothstein \cite{Goldberger:2004jt}. Instead,
we will use an alternative parametrization based on a temporal
Kaluza-Klein decomposition, as suggested by KS \cite{Kol:2007rx}. In
terms of the KS variables, the metric reads
\begin{equation} \label{eqn:KSmetric}
g_{\mu\nu}= \Bigg(
   \begin{array}{*{2}{c}}
    e^{2 \phi/m_{Pl}} & - e^{2 \phi/m_{Pl}} A_j /m_{Pl} \\
    - e^{2 \phi/m_{Pl}} A_i /m_{Pl} & \ \ \ -e^{-2 \phi/m_{Pl}} \gamma_{ij} + e^{2 \phi/m_{Pl}} A_i A_j /m_{Pl}^2 \\
   \end{array}
   \Bigg),
\end{equation}
where $\gamma_{ij} = \left(\delta_{ij} + \sigma_{ij} /m_{Pl}
\right)$. In four dimensions, the metric excitations are now
described in terms of a scalar field $\phi$, a three-vector field
$A_i$, and a $3\times3$ symmetric tensor field $\sigma_{ij}$, all of
which have been normalized to have mass dimension 1. In the ground
state, $\left<\phi\right> = \left<A_i\right> =
\left<\sigma_{ij}\right> = 0$, and the metric reduces to Minkowski
space-time. In terms of the KS variables, the EH action has the
simple form
\begin{equation} \label{eq:ehKS}
 S_{EH} = - 2 m_{Pl}^2 \int d^4x \sqrt{\gamma} \left(- R[\gamma] - \frac{1}{4} e^{4 \phi} F_{ij} F^{ij} + 2 (\partial^i \phi)(\partial_i \phi) + \ldots
 \right),
\end{equation}
where $F_{ij}=\partial_{i}A_{j}-\partial_{j}A_{i}$ and the dots
denote terms with time derivatives. In Eq. (\ref{eq:ehKS}) all
spatial indices are lowered or raised with $\gamma_{ij}$ or
$\gamma^{ij}$, respectively. Some of the terms involving time
derivatives are needed for our calculation along with the expansion
of the gauge fixing action of Eq. (\ref{eqn:gf}), and their
contributions enter the Feynman rules given in Sec. \ref{sec:calc}.

Given the gravitational action Eqs. (\ref{eqn:EHact}) and
(\ref{eqn:gf}), the worldline action from Eq. (\ref{eqn:pp}), and
the KS metric parametrization from Eq. (\ref{eqn:KSmetric}), we can
proceed to derive Feynman rules for the gravitational self-couplings
and the couplings to the compact object worldlines. The procedure is
equivalent to the one in \cite{Goldberger:2004jt}, but is now
implemented in terms of the KS fields. For the gravitational
self-couplings and worldline couplings, there are infinitely many
terms in the respective actions. To avoid unnecessary calculation at
a given order, a method of systematic computation must be
established.

\subsection{Power counting and Feynman diagram topologies}\label{sec:powertopo}

The power-counting rules for our calculation are the same as those
in the original EFT setup developed in \cite{Goldberger:2004jt}.
However, since there are no radiation fields in our 2PN potential
calculation, we use a simplified procedure where we power count our
Feynman diagrams relative to the LO Newtonian potential. For a bound
state, the virial theorem relates the orbital velocity $v$ and
Newton's constant $G$ via, $v^2\sim Gm/r$, and as shown in
\cite{Goldberger:2004jt}, one can power count all contributions in
powers of the orbital velocity $v$. Keeping the virial theorem in
mind, we count powers of $G$ and powers of $v^2$ separately. For the
power counting we do not keep track of factors of mass $m$ and
separation $r$ associated to each $G$. These will be generated
appropriately in the calculation of the diagrams. In this scheme,
the diagram responsible for the Newtonian potential scales as
$\mathcal O(G v^0)$. For the 2PN potential calculation, we need to
include all diagrams which scale as $\mathcal O(G v^4)$, $\mathcal
O(G^2 v^2)$, and $\mathcal O(G^3v^0)$.

To determine the relevant Feynman diagrams at 2PN, the first stage
is to generate all relevant diagram topologies. To do this, one
first counts in powers of $G$, since all interaction vertices will
scale with a power of $G$ (recall $m_{Pl}\sim G^{-1/2}$). Powers of
velocity are inserted later, when all relevant diagram topologies
have been established. There are two rules to consider when counting
a topology in factors of $G$. Firstly, when there are $n$ gravitons
attached to a worldline, this component receives a factor of
$G^{n/2}$. Secondly, each $n$-graviton self-interaction vertex
receives a factor of $G^{n/2-1}$. Note that a propagator does not
receive any factors. Using these rules any diagram topology can be
counted in powers of $G$.

\begin{figure}[tbp]
\centering \epsfig{file=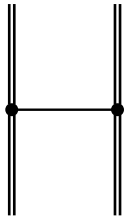} \caption{Order $G$ topology. The
single solid line denotes a generic graviton field, either $\phi$,
$A_i$, or $\sigma_{ij}$, and double solid lines denote the
worldlines of the binary constituents.} \label{fig:G1topo}
\end{figure}

\begin{figure}[tbp]
\centering \epsfig{file=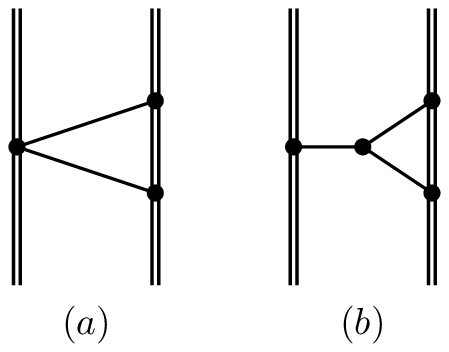} \caption{Topologies at
order $G^2$.} \label{fig:G2topo}
\end{figure}

\begin{figure}[tbp]
\centering \epsfig{file=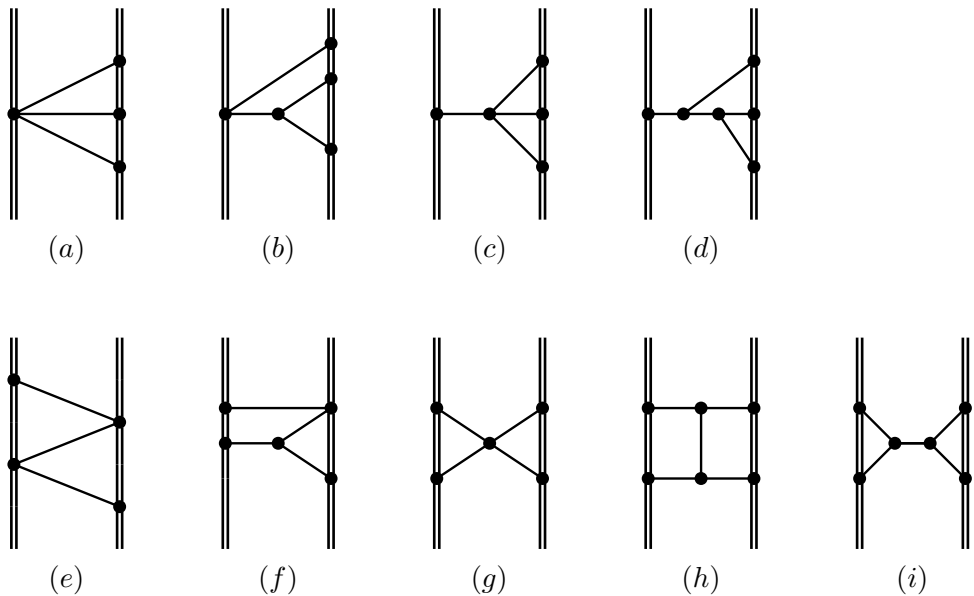} \caption{Topologies at order
$G^3$.} \label{fig:G3topo}
\end{figure}

If we now proceed to apply these rules, we can only have a single
topology at $\mathcal{O}(G)$, as shown in Fig. \ref{fig:G1topo}. All
topologies which simply renormalize the mass are omitted, as
discussed in \cite{Goldberger:2004jt}. Diagrams which involve a
graviton line starting and ending on the same worldline, without any
intermediate interaction, fall into this category. At the next order,
$\mathcal{O}(G^2)$, we have two topologies, as shown in Fig.
\ref{fig:G2topo}. Since the worldlines are static sources and do
not propagate, the diagrams in Fig. \ref{fig:G2topo} do not involve
any loops. Diagrams with gravitational loops yield quantum effects
and are therefore ignored, and since only massless fields propagate in NRGR,
the expansion in loops corresponds to an expansion in powers of $\hbar$
\cite{Holstein:2004dn}. At
$\mathcal{O}(G^3)$, the total number of possible topologies is nine,
as shown in Fig. \ref{fig:G3topo}. These diagrams are the relevant
topologies for the static component of the 2PN potential. In Figs.
\ref{fig:G2topo} and \ref{fig:G3topo}, we have
not drawn the topologies with the two worldlines interchanged,
although they will be required for our calculations.

Now that we have established all topologies relevant to the
calculation at 2PN, the powers of the orbital velocity $v$ must be
counted. There are two possible sources for factors of $v$ in our
calculations: (1) fields coupling to the worldlines, where the LO
couplings of $\phi$, $A_i$, and $\sigma_{ij}$ are $\mathcal O(v^0)$,
$\mathcal O(v^1)$, and $\mathcal O(v^2)$, respectively,  and (2)
time derivatives, where $\partial_0 \sim v/r$ for potential modes.
Time derivatives can arise from either purely gravitational
interaction vertices or from propagator insertions, where each
propagator insertion counts as $\mathcal O(v^2)$.

With all relevant topologies and the counting rules for $v$, the
final diagrams at 2PN can now be constructed. Recall that the 2PN
diagrams scale as either $\mathcal O(G v^4)$, $\mathcal O(G^2 v^2)$,
or $\mathcal O(G^3v^0)$. The diagrams are generated by first
populating the topologies with the three gravitational fields
$\phi$, $A_i$, and $\sigma_{ij}$ in all possible combinations. We
then count the factors of $v$ which result from the gravitational
fields coupling to the worldline, and any time derivatives acting on
internal vertices. Propagator insertions are also counted where
appropriate. The powers of $v$ from the worldline vertices are kept
to all orders in the calculations and only expanded to 2PN in the
final result. The diagrams relevant at 2PN which result from this
procedure are shown in Sec. \ref{sec:calc}, Figs. \ref{fig_2pnG1},
\ref{fig_2pnG2}, and \ref{fig_2pnG3}.

\subsection{Advantages of the Kol-Smolkin variables}\label{sec:KSadvan}

In the KS parametrization of Eq. (\ref{eqn:KSmetric}), we have
introduced three fields, $\phi$, $A_i$, and $\sigma_{ij}$, instead
of the standard $h_{\mu\nu}$ parametrization of Eq.
(\ref{eqn:hmunumetric}). It will be necessary to keep track of these
new fields, so one should ask what has been gained from a
calculational perspective by using the KS parametrization. This
question is answered by considering the static diagrams, i.e. those
with no velocity factors. First, consider the topology Fig.
\ref{fig:G2topo}$(b)$ which involves a three-graviton vertex. This
was used in \cite{Goldberger:2004jt} to construct the EIH potential
in the $h_{\mu\nu}$ variables. When the KS variables are employed,
it was shown that this topology does not contribute at 1PN order
\cite{Kol:2007bc}. We discuss this in more detail since it points us
to the source of the advantages of the KS variables.

When working in the $h_{\mu\nu}$ parametrization, one often
considers its components $h_{00}$, $h_{0i}$, and $h_{ij}$
separately, since their LO coupling to the worldline scale as
$\mathcal O(v^0)$, $\mathcal O(v^1)$, and $\mathcal O(v^2)$,
respectively. So in practice, in the $h_{\mu\nu}$ and KS
parametrizations, one must keep track of the same number of
components or fields. Returning to our discussion, if we consider
static diagrams, only the $h_{00}$ component can couple to the
worldline. Let us consider the static limit of the first topology
which arises with gravitational self-couplings, Fig.
\ref{fig:G2topo}$(b)$. With $h_{00}$ coupling to each worldline at
LO, without any propagator insertions, the only possible source of
powers of $v$ would be the three-graviton vertex. Naively, one would
isolate the action component with three powers of $h_{00}$, which
yields the 3-$h_{00}$ vertex. Interestingly, one finds that the
corresponding term in the action is proportional to $h_{00}
(\partial_0 h_{00})^2$, meaning that the 3-$h_{00}$ vertex has a
power of $v^2$ associated with it. So does this mean that the
topology in Fig. \ref{fig:G2topo}$(b)$ does not enter at 1PN in the
$h_{\mu\nu}$ parametrization? If this were true, then the EIH
calculation of \cite{Goldberger:2004jt} would be contradicted. The
reason why the above conclusion is wrong is the presence of mixing
of the $h_{\mu\nu}$ components.

To explain this point, consider the $h_{\mu\nu}$ graviton propagator
in harmonic gauge, which is given by $\left< T h_{\mu\nu} h_{\alpha
\beta} \right> \sim \left(\eta_{\mu \alpha} \eta_{\nu \beta} +
\eta_{\mu \beta} \eta_{\nu \alpha} - \eta_{\mu \nu} \eta_{\alpha
\beta}\right)$. Examining the tensor structure of the propagator,
one can see $h_{00}$ mixes with the trace of $h_{ij}$, since $\left<
T h_{00} h_{ij} \right> \sim \delta_{ij}$. Thus, a graviton which
starts as $h_{00}$ at the worldline coupling can be either $h_{00}$
or $h_{ii}$ at the three-graviton vertex. So instead of the
3-$h_{00}$ vertex, one has to use the three-point function $\left< T
h_{00} h_{00} h_{00}\right>$. This three-point function is obtained
by contracting the 3-$h_{\mu \nu}$ vertex with three graviton
propagators and setting all remaining free indices to $0$. One then
finds that there are components of the three-point function $\left<T
h_{00} h_{00} h_{00}\right>$ which do not involve time derivatives,
and so the topology Fig. \ref{fig:G2topo}$(b)$ is required at 1PN in
the $h_{\mu\nu}$ parametrization. Clearly, the topology Fig.
\ref{fig:G2topo}$(b)$ only survives at 1PN in the $h_{\mu\nu}$
parametrization because $h_{00}$ can mix with $h_{ii}$ as it
propagates.

The major advantage of the KS parametrization in harmonic gauge is
that it removes this kind of mixing, since the two-point functions
between the three fields are zero: $\langle T\phi A_i\rangle=\langle
T\phi\sigma_{jk}\rangle=\langle T A_i\sigma_{jk}\rangle=0$. For the
topology Fig. \ref{fig:G2topo}$(b)$ to contribute to the EIH
potential in the KS parametrization, all of the fields coupling to
the worldlines must be $\phi$'s, since only they couple at $\mathcal
O(v^0)$. Knowing that $\phi$ cannot mix with $A_i$ or $\sigma_{ij}$,
the 3-$\phi$ vertex must be used. When the gravitational Lagrangian
is examined, the only relevant term is proportional to
$\phi\dot{\phi}^2$, where the two time derivatives in this term
cause the 3-$\phi$ vertex to be of $\mathcal O(v^2)$ in the orbital
velocity. Therefore, the diagrams resulting from the topology Fig.
\ref{fig:G2topo}$(b)$ do not contribute until 2PN in the KS
parametrization.

This argument can be extended to the $n$-$\phi$ vertex $(n>2)$,
where the relevant term in the gravitational action is $\sim\exp(-
4\phi / m_{Pl})\dot{\phi}^2$. Any $n$-$\phi$ vertex is then suppressed by one
order in the PN expansion because there are always two time
derivatives. Thus, in the KS parametrization, the $\mathcal O(G^3)$
topologies Fig. \ref{fig:G3topo}$(b),(c),(f)$, and $(g)$ will not
enter at 2PN. This will reduce the calculational effort
considerably. Also, the fact that the topology Fig.
\ref{fig:G2topo}$(b)$ first enters at 2PN, rather than 1PN, means
that propagator insertions are not required for this topology at
2PN.

We have established that the KS variables in harmonic gauge
provide important calculational advantages over the
$h_{\mu\nu}$ parametrization used in \cite{Goldberger:2004jt} at 2PN
order. At 1PN, this advantage was modest, since only one diagram topology was
eliminated. At 2PN, we have found that four topologies at $\mathcal
O(G^3v^0)$ were removed, and one did not need propagator insertions
at $\mathcal O(G^2v^2)$ on the topology Fig. \ref{fig:G2topo}$(b)$.
Clearly as we proceed to higher order in the PN expansion, the
advantages of the KS variables become more important.

\section{Calculation of Feynman diagrams} \label{sec:calc}

In this section, we present the calculation of the Feynman diagrams
which are required to determine the 2PN interaction potential
between the binary constituents. Using the Feynman rules presented
below, each diagram equals $- i \int dt \, V$ from which we extract
its contribution to the potential $V$. The symmetry factor for each
diagram is computed in the usual manner \cite{Peskin:1995ev}, but
one needs to account for the nonpropagating classical sources. As
discussed in Sec. \ref{sec:powertopo}, there are three classes of
diagrams with the following power counting: $\mathcal O(G^3v^0)$,
$\mathcal O(G^2 v^2)$, and $\mathcal O(G v^4)$. We discuss how to
evaluate each in turn, starting with the simplest at $\mathcal O(G
v^4)$, and finishing with the more complex diagrams at $\mathcal
O(G^3v^0)$.

When calculating the diagrams, we work with the spatial Fourier
transform for the gravitational potential modes. For example, for
the $\phi$ field, we will work in terms of $\phi_\mathbf k (t)$,
which is the spatial Fourier transform of the coordinate space field
$\phi(t,\mathbf x) = \int_{\mathbf k} e^{i \mathbf{k \cdot x}}
\phi_\mathbf k (t)$, where $\int_{\mathbf k} = \int \frac{d^3
\mathbf k}{(2 \pi)^3}$. We also make clear that after expanding the
metric in the KS fields $\phi$, $A_i$, and $\sigma_{ij}$, we do not
distinguish between upper and lower spatial indices; they are
lowered and raised with $\delta_{ij}$ and $\delta^{ij}$,
respectively. Lastly, since we are interested in the long range
potential only, we drop irrelevant contact terms, such as
$V\sim\delta(r)$, wherever they appear in the calculations.

\subsection{Order $Gv^4$ diagrams}\label{sec:Gv4}

\begin{figure}
\begin{center}
\epsfig{file=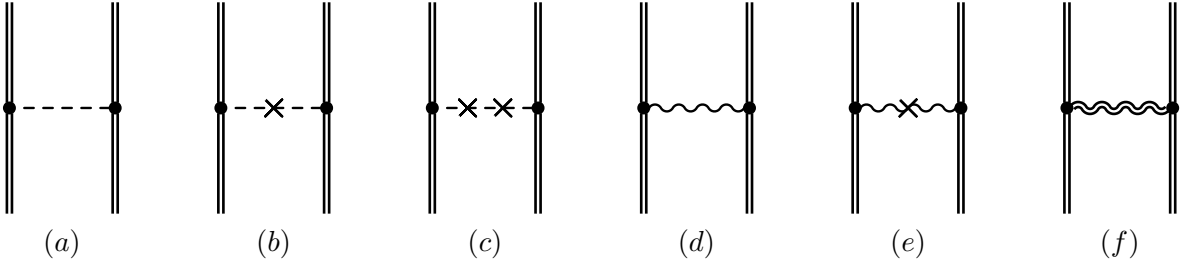} \caption{Order $Gv^4$ diagrams at 2PN. Here
the dashed, wavy, and double wavy lines, represent the $\phi$,
$A_i$, and $\sigma_{ij}$ fields, respectively. A cross denotes a
propagator insertion.} \label{fig_2pnG1}
\end{center}
\end{figure}

As shown in Fig. \ref{fig_2pnG1}, there are six diagrams to be
evaluated at $\mathcal O(Gv^4)$. The contributions from the diagrams
$(a)$, $(d)$, and $(f)$ are the easiest of all diagrams to compute
at 2PN, since we only need to compute simple one-graviton exchange
diagrams.

The Feynman rules for couplings to the worldlines are derived from
the point particle action in Eq. (\ref{eqn:pp}), which is expanded
in the gravitational fields to the required order for a given
Feynman rule. The Feynman rules for a $\phi$, $A_{i}$, and
$\sigma_{ij}$ coupled to the worldline are
\begin{align}
\label{eqn:worldphi}  \parbox{8mm}{\includegraphics{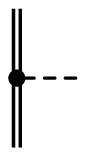}}
 & = - \frac{i m}{m_{Pl}} \! \int \! dt \! \int_{\mathbf k} \! e^{i \mathbf{k \cdot x}} \frac{1 \! + \! \mathbf v^2}{\sqrt{1 \! - \! \mathbf v^2}} \\ {} \notag\\
\label{eqn:worldA}   \parbox{8mm}{\includegraphics{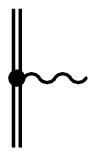}}
 & = \frac{i m}{m_{Pl}} \! \int \! dt \! \int_{\mathbf k} \! e^{i \mathbf{k \cdot x}} \frac{\mathbf v_i}{\sqrt{1 \! - \! \mathbf v^2}} \\ {} \notag\\
\label{eqn:worldsigma}   \parbox{8mm}{\includegraphics{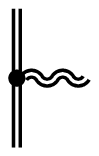}}
 & = \frac{i m}{2 m_{Pl}} \! \int \! dt \! \int_{\mathbf k} \! e^{i \mathbf{k \cdot x}} \frac{\mathbf v_i \mathbf v_j}{\sqrt{1 \! - \! \mathbf
v^2}}.
\end{align}
The double solid lines are the worldlines, and the dashed, wavy, and
double wavy lines represent the $\phi$, $A_i$, and $\sigma_{ij}$
fields, respectively. Note that these rules are exact in $v$. We
will expand in $v$ at the end of the calculation to isolate the
required terms at 2PN.

To compute diagrams $(a)$, $(d)$, and $(f)$, we now need the
propagators for each gravitational field. From the purely
gravitational action, the sum of Eqs. (\ref{eqn:EHact}) and
(\ref{eqn:gf}), the propagators can be derived. These potential
propagators are nonrelativistic and instantaneous, and for the
$\phi$, $A_{i}$, and $\sigma_{ij}$ gravitational fields we have,
respectively,
\begin{align}
 \label{eqn:propphi} \left<T \phi_{\mathbf p}(t_a) \phi_{\mathbf q}(t_b)\right> & = - \frac{1}{8} (2 \pi)^3 \delta^3(\mathbf p + \mathbf q) \frac{i}{\mathbf p^2} \delta(t_a - t_b) \\
 \label{eqn:propA} \left<T A^i_{\mathbf p}(t_a) A^j_{\mathbf q}(t_b)\right> & = \frac{1}{2} (2 \pi)^3 \delta^3(\mathbf p + \mathbf q) \frac{i \, \delta^{ij}}{\mathbf p^2} \delta(t_a - t_b) \\
\label{eqn:propsig} \left<T \sigma_{\mathbf p}^{ij}(t_a)
\sigma_{\mathbf q}^{kl}(t_b)\right> & = - (2 \pi)^3 \delta^3(\mathbf
p + \mathbf q) \frac{i P^{ij,kl}}{\mathbf p^2} \delta(t_a - t_b),
\end{align}
where $P^{ij,kl} = \frac{1}{2}(\delta^{ik} \delta^{jl} + \delta^{il}
\delta^{jk} - 2 \delta^{ij} \delta^{kl})$.

The three single-graviton exchange diagrams are easily computed with the
above propagators and worldline couplings. At $\mathcal O(Gv^4)$,
the only integral in $\mathbf k$ which arises is a Fourier transform,
which is evaluated using the $d$-dimensional formula
\begin{equation}\label{eq_FTint}
 \int \frac{d^d \mathbf k}{(2 \pi)^d} \frac{1}{(\mathbf k^2)^\alpha} e^{i \mathbf k \cdot \mathbf r} = \frac{1}{(4 \pi)^{d/2}} \frac{\Gamma(d/2 - \alpha)}{\Gamma(\alpha)} \left(\frac{\mathbf r^2}{4}\right)^{\alpha-d/2}.
\end{equation}
Doing the Fourier transform, and extracting the 2PN component, gives
the following potential contributions:
\begin{align}
 V^{(a)} = & - \frac{G m_1 m_2}{r} \left(\frac{7}{8} \mathbf v_1^4 + \frac{9}{4} \mathbf v_1^2 \mathbf v_2^2 + \frac{7}{8} \mathbf v_2^4 \right) \\
 V^{(d)} = & - \frac{G m_1 m_2}{r} \left(- 2 \mathbf v_1 \cdot \mathbf v_2 \left(\mathbf v_1^2 + \mathbf v_2^2 \right)\right) \\
 V^{(f)} = & - \frac{G m_1 m_2}{r} \left(2 (\mathbf v_1 \cdot \mathbf v_2)^2 - 2 \mathbf v_1^2 \mathbf v_2^2\right),
\end{align}
where the notation we use here and subsequently is $\mathbf r \equiv
\mathbf x_1(t) - \mathbf x_2(t)$, $r \equiv \left|\mathbf r\right|$,
and $\mathbf n \equiv \mathbf r / r$. The labels 1 and 2 are used
for the left and right worldlines, respectively. Although $\mathbf
r$, $\mathbf v_1$, etc., depend on $t$, we have suppressed this
dependence above. We will continue to do this when convenient.

Diagrams $(b)$, $(c)$, and $(e)$ all involve propagator insertions.
These insertions account for the corrections to the instantaneous
nature of the nonrelativistic potential propagators in Eqs.
(\ref{eqn:propphi}) - (\ref{eqn:propsig}). These are included
systematically as a perturbation, where each propagator insertion is
suppressed with respect to the propagator by a power of $v^2$. The
insertions are obtained from the terms in the action which are
quadratic in the fields and which involve time derivatives. The
required propagator insertions at 2PN are derived to be
\begin{align}
\label{eqn:insert1phi}\parbox{22mm}{\includegraphics{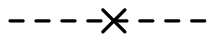}}
& =-\frac{1}{8}(2\pi)^3\delta^3(\mathbf p +\mathbf q)\frac{i}{\mathbf p^4}\partial_{t_a}\partial_{t_b}\delta(t_a-t_b) {} \\
\label{eqn:insert2phi}\parbox{22mm}{\includegraphics{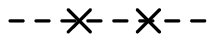}}
& =-\frac{1}{8}(2\pi)^3\delta^3(\mathbf p +\mathbf q)\frac{i}{\mathbf p^6}\partial_{t_a}^2\partial_{t_b}^2\delta(t_a-t_b) {} \\
\label{eqn:insert1A}\parbox{22mm}{\includegraphics{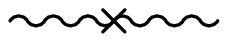}}
&=\frac{1}{2}(2\pi)^3\delta^3(\mathbf p+\mathbf
q)\frac{i\,\delta^{ij}}{\mathbf
p^4}\partial_{t_a}\partial_{t_b}\delta(t_a-t_b){},
\end{align}
where one cross denotes a single propagator insertion and two
crosses denote two propagator insertions. Here, we have chosen the
simplest symmetric form of the time derivatives; however, other
choices are possible. Note that any other choice of time derivatives
yields potentials which can be related to potentials computed with
the simplest symmetric choice by a total time derivative, and are
thus physically equivalent.

To demonstrate how the diagrams at $\mathcal O(Gv^4)$ are computed,
we show how to calculate diagram $(b)$ in some detail. The first
step is to form the expression of the diagram. This
is achieved by coupling Eq. (\ref{eqn:insert1phi}) to two copies of
Eq. (\ref{eqn:worldphi}) for worldlines 1 and 2. The diagram is
given as follows, where we have done the $\mathbf q$ integration to
eliminate $\delta^3\left(\mathbf p+\mathbf q\right)$,
\begin{equation}\label{eqn:ampGv4_1}
     -i\int dt \, V =\frac{i m_1 m_2}{8 m_{Pl}^2}\int dt_a dt_b\int_{\mathbf p}\frac{e^{i \mathbf p \cdot \left(\mathbf x_1\left(t_a\right)-\mathbf x_2\left(t_b\right)\right)}}{\mathbf
    p^4}\frac{(1+\mathbf v_1(t_a)^2)(1+\mathbf v_2(t_b)^2)}{(1-\mathbf v_1(t_a)^2)^{1/2}(1-\mathbf v_2(t_b)^2)^{1/2}}\partial_{t_a} \partial_{t_b} \delta(t_a -
t_b).
\end{equation}
Care must be taken when labeling the time for each worldline Feynman
rule, and here we have used $t_a$ and $t_b$. Since each worldline
interaction is a different Feynman rule, it must therefore have a different
dummy variable. Before explicitly using the time derivatives, we
employ the Fourier integral from Eq. (\ref{eq_FTint}) to compute
the remaining momentum integral. Now, an integration
by parts on both time derivatives acting on $\delta(t_a - t_b)$ is
required, which gives,
\begin{equation}\label{eqn:ampGv4_2}
     -i\int dt \, V =\frac{-i G m_1 m_2}{2}\int dt_a dt_b \delta(t_a - t_b)\left\{ \partial_{t_a}
\partial_{t_b}\left[
    \left|\mathbf x_1(t_a)-\mathbf x_2(t_b)\right|\frac{(1+\mathbf v_1(t_a)^2)(1+\mathbf v_2(t_b)^2)}{(1-\mathbf v_1(t_a)^2)^{1/2}(1-\mathbf
v_2(t_b)^2)^{1/2}}\right]\right\}.
\end{equation}
At this stage, the time derivatives must be allowed to act on the
square bracket. Only after this point can the $\delta(t_a - t_b)$ be
used. It is clear that acceleration-dependent terms will arise. The
potential contribution we obtain is exact to all orders in the PN
expansion. Since we are only interested in the 2PN potential, we
extract this piece from the result. For this purpose, we note that
accelerations are power-counted as $a \sim v^2/r$. The final
contribution, $V^{(b)}$, is given below, along with the other
one-graviton exchange diagrams with propagator insertions at
$\mathcal O(Gv^4)$,
\begin{align}
V^{(b)} = & - \frac{G m_1 m_2}{r} \left(\frac{3}{4} \left(\mathbf v_1 \cdot \mathbf v_2 - \mathbf n \cdot \mathbf v_1 \mathbf n \cdot \mathbf v_2\right) \left(\mathbf v_1^2 + \mathbf v_2^2\right) \right) \notag \\
                 & - G m_1 m_2 \left(\frac{3}{2} \mathbf a_1 \cdot \mathbf v_1 \mathbf n \cdot\mathbf v_2 - \frac{3}{2} \mathbf a_2 \cdot \mathbf v_2 \mathbf n \cdot\mathbf v_1\right)\\
V^{(c)} = & - \frac{G m_1 m_2}{r} \left(\frac{1}{8} \mathbf v_1^2 \mathbf v_2^2 + \frac{1}{4} (\mathbf v_1 \cdot \mathbf v_2)^2 + \frac{3}{8} (\mathbf n \cdot \mathbf v_1)^2(\mathbf n \cdot \mathbf v_2)^2\right. \notag \\
                 & \left. {} \hspace*{48pt} - \frac{1}{2} \mathbf v_1 \cdot \mathbf v_2 \, \mathbf n \cdot \mathbf v_1 \mathbf n \cdot \mathbf v_2 - \frac{1}{8} \mathbf v_1^2 (\mathbf n \cdot \mathbf v_2)^2 - \frac{1}{8} \mathbf v_2^2 (\mathbf n \cdot \mathbf v_1)^2 \right) \notag \\
                 & - G m_1 m_2 \left(\frac{1}{8} \mathbf a_1 \cdot \mathbf n \left(\mathbf v_2^2 - (\mathbf n \cdot \mathbf v_2)^2\right) - \frac{1}{8} \mathbf a_2 \cdot \mathbf n \left(\mathbf v_1^2 - (\mathbf n \cdot \mathbf v_1)^2\right) \right. \notag \\
                 & \left. {} \hspace*{44pt} + \frac{1}{4} \mathbf a_1 \cdot \mathbf v_2 \mathbf n \cdot \mathbf v_2 - \frac{1}{4} \mathbf a_2 \cdot \mathbf v_1 \mathbf n \cdot \mathbf v_1\right) \notag \\
                 & - G m_1 m_2 r \left(- \frac{1}{8} \mathbf a_1 \cdot \mathbf a_2 - \frac{1}{8} \mathbf a_1 \cdot \mathbf n \mathbf a_2 \cdot \mathbf n \right) \\
V^{(e)} = & - \frac{G m_1 m_2}{r} \left(- 2 (\mathbf v_1 \cdot \mathbf v_2)^2 + 2 \mathbf v_1 \cdot \mathbf v_2 \mathbf n \cdot \mathbf v_1 \mathbf n \cdot \mathbf v_2 \right) \notag \\
                 & - G m_1 m_2 \left(-2 \mathbf a_1 \cdot \mathbf v_2 \mathbf n \cdot \mathbf v_2 + 2 \mathbf a_2 \cdot \mathbf v_1 \mathbf n \cdot \mathbf v_1 \right) \notag \\
                 & - G m_1 m_2 r \left(2 \mathbf a_1 \cdot \mathbf a_2
\right).
\end{align}
Note that all the potential contributions $V^{(b)}$, $V^{(c)}$, and
$V^{(e)}$, contain acceleration-dependent terms. They are generated
from either a time derivative acting on a worldline velocity factor
or from two time derivatives acting on $\mathbf x_{1}$ or $\mathbf
x_{2}$.

\subsection{Order $G^2v^2$ diagrams}\label{sec:G2v2}

\begin{figure}
\begin{center}
\epsfig{file=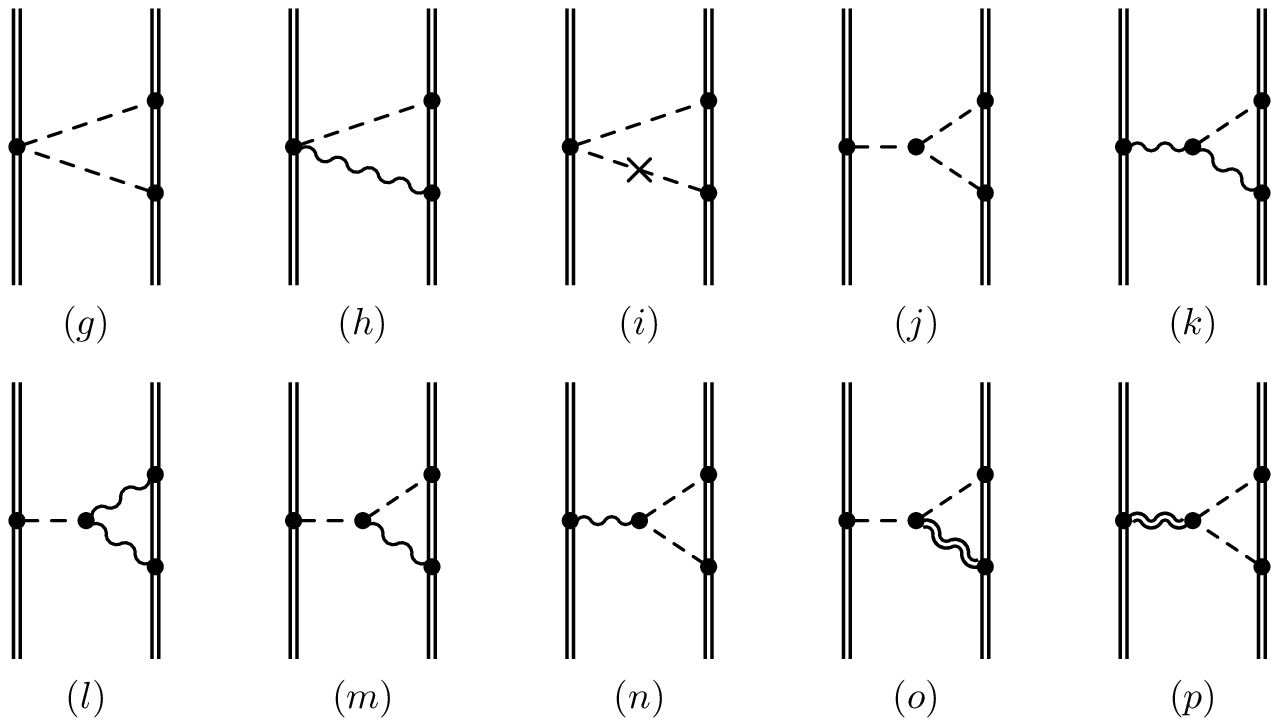} \caption{Order $G^2v^2$ diagrams at 2PN.}
\label{fig_2pnG2}
\end{center}
\end{figure}

The first three diagrams at $\mathcal O(G^2 v^2)$, $(g)$, $(h)$, and $(i)$ of Fig.
\ref{fig_2pnG2}, are obtained by modifying the
$\mathcal O(G)$ topology through the addition of one extra
gravitational leg connecting the two worldlines. As we will see,
such diagrams, where a topology of lower order in $G$ is augmented
by an additional leg between the worldlines, are particularly simple
to compute because their expressions factorize. To compute these
diagrams we need two additional Feynman rules for the
worldline couplings. The first has two $\phi$ fields coupling to the
worldline, and the second has a $\phi$ and an $A_{i}$ coupling to
the worldline. These are respectively,
\begin{align}
\label{eqn:worldphiphi}  \parbox{8mm}{\includegraphics{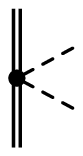}}
 & = - \frac{i m}{m_{Pl}^2} \! \int \! dt \! \int_{\mathbf k, \mathbf q} \! \! \! \! e^{i \mathbf{(k+q) \cdot x}} \frac{1 \! - \! 6 \mathbf v^2 \! + \! \mathbf v^4}{\left(1 \! - \! \mathbf v^2\right)^{3/2}}  \\ {}\notag \\
\label{eqn:worldphiA}  \parbox{8mm}{\includegraphics{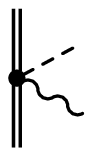}}
 & = \frac{i m}{m_{Pl}^2} \! \int \! dt \! \int_{\mathbf k, \mathbf q} \! \! \! \! e^{i \mathbf{(k+q) \cdot x}} \frac{(1 \! - \! 3 \mathbf v^2)\mathbf v_i}{\left(1 \! - \! \mathbf v^2\right)^{3/2}}. \\
{}\notag
\end{align}
These rules are again exact in the orbital velocity $v$.

We demonstrate how to calculate the three diagrams $(g)$, $(h)$, and
$(i)$, by considering diagram $(h)$. One can write this diagram by
using Eqs. (\ref{eqn:worldphi}), (\ref{eqn:worldA}), and
(\ref{eqn:worldphiA}) for the worldline couplings, and Eqs.
(\ref{eqn:propphi}) and (\ref{eqn:propA}) for the propagators. The
diagram is
\begin{align}\notag
     -i\int dt\,V=&\frac{i m_1 m_2^2}{16 m_{Pl}^4}\int dt_a dt_b dt_c \delta(t_a - t_b) \delta(t_a - t_c) \int_{\mathbf k,\mathbf l, \mathbf p,\mathbf q} e^{i \left[(\mathbf k+\mathbf p) \cdot \mathbf x_1(t_a)+\mathbf l \cdot \mathbf x_2(t_b)+\mathbf q \cdot \mathbf
x_2(t_c)\right]}\\\label{eqn:ampG2v2_1}
&\times\frac{(2\pi)^3\delta^3\left(\mathbf k+\mathbf
l\right)(2\pi)^3\delta^3\left(\mathbf p+\mathbf q\right)}{\mathbf
    k^2 \mathbf p^2}\frac{\mathbf v_1(t_a)\cdot\mathbf v_2(t_c)(1-3 \mathbf v_1(t_a)^2)(1+\mathbf v_2(t_b)^2)}{(1-\mathbf v_1(t_a)^2)^{3/2}(1-\mathbf v_2(t_b)^2)^{1/2}(1-\mathbf
v_2(t_c)^2)^{1/2}}.
\end{align}
The symmetry factor of this diagram is $1$. Since there are no time
derivatives in this expression, the time delta functions can be used
immediately. We can also use the three-momentum delta functions;
doing so leads to
\begin{equation}\label{eqn:ampG2v2_2}
     -i\int dt\,V=\frac{i m_1 m_2^2}{16 m_{Pl}^4}\int dt \int_{\mathbf k,\mathbf p} \frac{e^{i \mathbf k \cdot\mathbf r}}{\mathbf k^2}\frac{e^{i \mathbf p \cdot\mathbf r}}{\mathbf p^2}\frac{\mathbf
v_1\cdot\mathbf v_2(1-3 \mathbf v_1^2)(1+\mathbf v_2^2)}{(1-\mathbf
v_1^2)^{3/2}(1-\mathbf v_2^2)}.
\end{equation}
Looking at this equation, it is clear that the two momentum
integrations factorize into two Fourier transforms. They are seen to
be the same in this diagram, and we can simply use the Fourier
integral, Eq. (\ref{eq_FTint}). Our final result for the exact
potential contribution from diagram $(h)$ to all orders in the PN
expansion, is then
\begin{equation}\label{eqn:ampG2v2_3}
    V^{(h)}_{\textrm{Exact}}=\frac{-4 G^2 m_1 m_2^2}{r^2}\frac{\mathbf v_1\cdot\mathbf
v_2(1-3 \mathbf v_1^2)(1+\mathbf v_2^2)}{(1-\mathbf
v_1^2)^{3/2}(1-\mathbf v_2^2)}.
\end{equation}

The calculation of the other two diagrams with this topology,
diagrams $(g)$ and $(i)$, proceed similarly. However, when
propagator insertions are present, as in diagram $(i)$, the time
delta functions cannot be used immediately. One must first integrate
by parts, and act with the time derivatives, before using the time
delta functions. The procedure is analogous to diagram $(b)$ in Sec.
\ref{sec:Gv4}. At 2PN, we have the following contributions to the
potential:
\begin{align}
 V^{(g)} = & \frac{G^2 m_1 m_2^2}{r^2} \left(- \frac{9}{4} \mathbf v_1^2 + \frac{3}{2} \mathbf v_2^2 \right)  \\
 V^{(h)} = & \frac{G^2 m_1 m_2^2}{r^2} \left(- 4 \mathbf v_1 \cdot \mathbf v_2 \right)  \\
 V^{(i)} = & \frac{G^2 m_1 m_2^2}{r^2} \left(\frac{1}{2} (\mathbf n \cdot \mathbf v_2)^2 +
\frac{1}{2} \mathbf v_1 \cdot \mathbf v_2 - \mathbf n \cdot \mathbf v_1 \mathbf n \cdot \mathbf v_2 \right).
\end{align}
Although we do not display the diagrams where the two worldlines are
interchanged, they must be accounted for in the final 2PN
Lagrangian.

We now proceed to the diagrams where the topology is that of Fig.
\ref{fig:G2topo}(b), where a three-graviton vertex mediates the
interaction. First, the vertices for the gravitational
self-interactions must be derived, and this is done in a
straightforward way from the sum of the Einstein-Hilbert action, Eq.
(\ref{eqn:EHact}), and the gauge fixing action, Eq. (\ref{eqn:gf}).
At 2PN, the KS parametrization simplifies the vertex derivations,
since we need at most one leg to be the tensor field $\sigma_{ij}$.
Once the appropriate terms in the action are established via
power-counting factors of $v$, the Feynman rule for the vertex is
computed. Then each leg of a given vertex is multiplied by the
appropriate propagator to obtain the three-point function. For
diagrams $(j)$, $(k)$, $(l)$, $(m)$, $(n)$, $(o)$, and $(p)$ of Fig.
\ref{fig_2pnG2}, we use the following three-point functions:
\begin{align}
 \left<T \phi_a \phi_b \phi_c \right> & = \frac{1}{16 m_{Pl}} \Bigg\{\Big(\partial_{t_a} \delta(t_a - t_c)\Big) \Big(\partial_{t_b} \delta(t_b - t_c)\Big)
                                                                  + \Big(\partial_{t_a} \delta(t_a - t_b)\Big) \Big(\partial_{t_c} \delta(t_c - t_b)\Big)\notag \\
 \label{eqn:3ptppp} & \hspace*{48pt}             + \Big(\partial_{t_b} \delta(t_b - t_a)\Big) \Big(\partial_{t_c} \delta(t_c - t_a)\Big)\Bigg\} \times \frac{1}{\mathbf k_a^2 \mathbf k_b^2 \mathbf k_c^2} \, (2 \pi)^3 \, \delta^3 (\mathbf k_a + \mathbf k_b + \mathbf k_c) \\
 \label{eqn:3ptppA}\left<T \phi_a \phi_b A_c^i \right> & = - \frac{i}{16 m_{Pl}} \Bigg\{\Big(\mathbf k_b^i \partial_{t_a} +\mathbf k_a^i \partial_{t_b}\Big) \delta(t_a - t_c) \delta(t_b - t_c) \Bigg\} \times \frac{1}{\mathbf k_a^2 \mathbf k_b^2 \mathbf k_c^2} \, (2 \pi)^3 \, \delta^3 (\mathbf k_a + \mathbf k_b + \mathbf k_c) \\
 \label{eqn:3ptpAA}\left<T \phi_a A_b^i A_c^j \right> & = \frac{1}{4 m_{Pl}} \, \delta(t_a - t_c) \delta(t_b - t_c) \times \frac{\mathbf k_b^i \mathbf k_c^j - \mathbf k_c^i \mathbf k_b^j + \mathbf k_b \cdot \mathbf k_c \delta^{ij}}{\mathbf k_a^2 \mathbf k_b^2 \mathbf k_c^2} \, (2 \pi)^3 \, \delta^3 (\mathbf k_a + \mathbf k_b + \mathbf k_c) \\
 \label{eqn:3ptppsig}\left<T \phi_a \phi_b \sigma_c^{ij} \right> & = \frac{1}{16 m_{Pl}} \, \delta(t_a - t_c) \delta(t_b - t_c) \times \frac{\mathbf k_a^i \mathbf k_b^j +\mathbf k_b^i \mathbf k_a^j}{\mathbf k_a^2 \mathbf k_b^2 \mathbf k_c^2} \, (2 \pi)^3 \, \delta^3 (\mathbf k_a + \mathbf k_b + \mathbf
k_c),
\end{align}
where the lower index on a field denotes its dependence on $t$ and
$\mathbf k$.

Because of the nonlinear structure of the diagrams $(j)$ through
$(p)$, their contributions do not factorize, and we have to
calculate an integral which corresponds to a one-loop Feynman
integral. To demonstrate the calculation of this class of diagrams,
we consider diagram $(k)$. This diagram uses the three-point
function $\left<T \phi_a A_b^i A_c^j \right>$ from Eq.
(\ref{eqn:3ptpAA}). The diagram is formed by coupling Eq.
(\ref{eqn:3ptpAA}) to the worldlines using Eqs. (\ref{eqn:worldphi})
and (\ref{eqn:worldA}) and the symmetry factor of this diagram is 1.
Given there are no time derivatives, we are free to do two of the
time integrations immediately. We also do the integration over the
three-momentum which labels the $\phi$ leg. The result of these
operations is
\begin{equation}\label{eqn:ampG2v2_1a}
-i\int dt \, V = \frac{i m_1 m_2^2}{4 m_{Pl}^4} \int dt
\frac{(1+\mathbf v_2^2)\mathbf v_1^i\mathbf v_2^j}{(1-\mathbf
v_1^2)^{1/2}(1-\mathbf v_2^2)}I^{ij}(\mathbf r),
\end{equation}
where the integral $I^{ij}(\mathbf r)$ is given by
\begin{equation}\label{eqn:ampG2v2_1b}
I^{ij}(\mathbf r)=\int_{\mathbf p,\mathbf q} e^{i \mathbf p\cdot
\mathbf r} \frac{\mathbf p^i \mathbf q^j - \mathbf q^i \mathbf p^j +
\mathbf p \cdot \mathbf q \delta^{ij}}{ \mathbf p^2 \mathbf q^2
(\mathbf p + \mathbf q)^2} = \int_\mathbf p e^{i \mathbf p\cdot
\mathbf r} \frac{\mathbf p^i \delta^{ja} - \mathbf p^j \delta^{ia} +
\mathbf p^a \delta^{ij}}{\mathbf p^2} \int_\mathbf q \frac{\mathbf
q^a}{\mathbf q^2 (\mathbf q + \mathbf p)^2}.
\end{equation}
The next step is to evaluate the momentum integral, $I^{ij}(\mathbf
r)$. From Eq. (\ref{eqn:ampG2v2_1b}) it is apparent that
$I^{ij}(\mathbf r)$ is calculated by first computing the $\mathbf
q$-integration which corresponds to a one-loop Feynman integral and
subsequently performing a Fourier integration over $\mathbf p$. The
Feynman integral is of vector nature, and we reduce it to a scalar
integral in the usual way by noting that its expression must be
proportional to $\mathbf p^a$. The scalar integral is then computed
in dimensional regularization using the $d$-dimensional master
integral
\begin{equation}\label{eq_FMint1}
\int \frac{d^d \mathbf k}{(2 \pi)^d} \frac{1}{\left[(\mathbf k + \mathbf
p)^2 \right]^{n_1} \left[\mathbf k^2 \right]^{n_2}} = \frac{1}{(4
\pi)^{d/2}} \frac{\Gamma (n_1 + n_2 - d/2)}{\Gamma (n_1)
\Gamma(n_2)} \frac{\Gamma(d/2-n_1)
\Gamma(d/2-n_2)}{\Gamma(d-n_1-n_2)} (\mathbf p^2)^{d/2-n_1-n_2}.
\end{equation}
The resulting expression for the integral $I^{ij}(\mathbf r)$ then
becomes
\begin{equation}
I^{ij}(\mathbf r) = - \frac{\delta^{ij}}{16} \int_\mathbf p e^{i
\mathbf p\cdot \mathbf r} \frac{1}{|\mathbf p|},
\end{equation}
and the remaining Fourier transform integral is performed with Eq. (\ref{eq_FTint}).
Putting all terms back together gives
the exact potential for diagram $(k)$,
\begin{equation}\label{eqn:ampG2v2_1exact}
V^{(k)}_{\textrm{Exact}}=\frac{G^2 m_1 m_2^2}{r^2} \frac{8 \mathbf
v_1 \cdot \mathbf v_2(1+\mathbf v_2^2)}{(1-\mathbf
v_1^2)^{1/2}(1-\mathbf v_2^2)}.
\end{equation}
The 2PN piece of this potential is extracted below.

The remaining diagrams from $(j)$ through $(p)$ are all evaluated
using the same methods as explained for diagram $(k)$. The only
complication are the time derivatives which occur in the three-point
functions $\left<T \phi_a \phi_b \phi_c \right>$ and $\left<T \phi_a
\phi_b A_c^i \right>$, Eqs. (\ref{eqn:3ptppp}) and
(\ref{eqn:3ptppA}), respectively. So when calculating the diagrams
$(j)$, $(m)$, and $(n)$, it is necessary to integrate by parts as
explained in Sec. \ref{sec:Gv4}. We use  Eq. (\ref{eq_FMint1}) (and
its vector or tensor extensions) for any one-loop integrals, and Eq.
(\ref{eq_FTint}) for the Fourier transforms. Additionally, for some
diagrams we need vector and tensor Fourier integrals which are
obtained from the scalar Fourier integral in Eq. (\ref{eq_FTint}) by
taking derivatives with respect to $\mathbf r$. At $\mathcal O
(G^2v^2)$, the contributions from the individual diagrams with the
topology Fig. \ref{fig:G2topo}(b) are
\begin{align}
 V^{(j)} = & \frac{G^2 m_1 m_2^2}{r^2} \left(\frac{1}{2} \mathbf v_2^2 - \frac{1}{2} (\mathbf n \cdot \mathbf v_2)^2 - 2 \mathbf v_1 \cdot \mathbf v_2 + 4 \mathbf n \cdot \mathbf v_1 \mathbf n \cdot \mathbf v_2 \right)  \\
 V^{(k)} = & \frac{G^2 m_1 m_2^2}{r^2} \left(8 \mathbf v_1 \cdot \mathbf v_2\right)  \\
 V^{(l)} = & \frac{G^2 m_1 m_2^2}{r^2} \left(- 4 \mathbf v_2^2 \right)  \\
 V^{(m)} = & \frac{G^2 m_1 m_2^2}{r^2} \left(2 \mathbf v_2^2 - 4 (\mathbf n \cdot \mathbf v_2)^2 + 2 \mathbf v_1 \cdot \mathbf v_2 - 4 \mathbf n \cdot \mathbf v_1 \mathbf n \cdot \mathbf v_2\right)  \\
 V^{(n)} = & \frac{G^2 m_1 m_2^2}{r^2} \left(- \mathbf v_1 \cdot \mathbf v_2 + \mathbf n \cdot \mathbf v_1 \mathbf n \cdot \mathbf v_2\right)  \\
 V^{(o)} = & \frac{G^2 m_1 m_2^2}{r^2} \left(- 2 \mathbf v_2^2 + 4 (\mathbf n \cdot \mathbf v_2)^2\right)  \\
 V^{(p)} = & \frac{G^2 m_1 m_2^2}{r^2} \left(\frac{1}{2} \mathbf v_1^2 - \frac{1}{2} (\mathbf n \cdot \mathbf
 v_1)^2\right).
\end{align}
All of the above contributions with $(1 \leftrightarrow 2)$ must be
added to the final potential in order to account for the diagrams
with the two worldlines interchanged.

\subsection{Order $G^3v^0$ diagrams}

\begin{figure}
\begin{center}
\epsfig{file=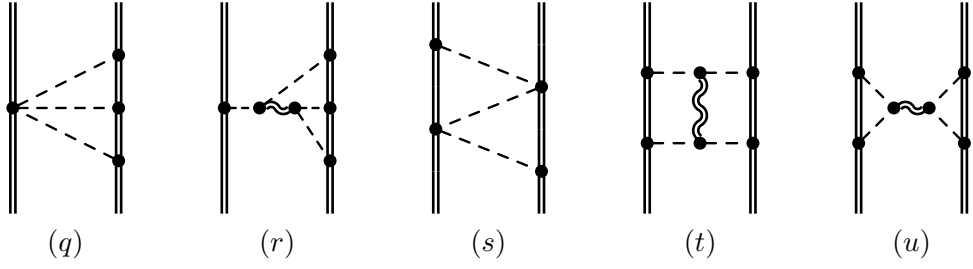} \caption{Order $G^3 v^0$ diagrams at 2PN.}
\label{fig_2pnG3}
\end{center}
\end{figure}

Finally, we progress onto the diagrams at $\mathcal O(G^3v^0)$. The
diagrams which must be evaluated are $(q)$, $(r)$, $(s)$, $(t)$, and
$(u)$, as shown in Fig. \ref{fig_2pnG3}. For diagram $(q)$ a new
worldline vertex will be required coupling three $\phi$ fields to
the worldline at one point. The Feynman rule is
\begin{align}
  \parbox{8mm}{\includegraphics{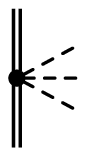}}
 & = - \frac{i m}{m_{Pl}^3} \! \int \! dt \! \int_{\mathbf k, \mathbf q, \mathbf p} \! \! \! \! \! \! \! \! e^{i \mathbf{(k+q+p) \cdot x}} \frac{1 \! + \! 11 \mathbf v^2 \! + \! 11 \mathbf v^4 \! + \! \mathbf v^6}{\left(1 \! - \! \mathbf v^2\right)^{5/2}},
\end{align}
which is exact to all orders in $v$. At this order, the
potential contribution at 2PN is static and there are no velocity
factors in the final potential contributions. Given this is the
case, we do not keep any velocity factors from the worldline
vertices when we discuss the calculations below.

The two simple diagrams at this order are $(q)$ and $(s)$. These
diagrams are easy because there are no internal vertices -- only
propagators and worldline couplings. The diagrams therefore factor
into simple Fourier transforms, which are evaluated using Eq.
(\ref{eq_FTint}). This is analogous to the evaluation of diagram
$(h)$ in Sec. \ref{sec:G2v2}. The symmetry factors are $1/6$ and $1$
for $(q)$ and $(s)$, respectively. Computing each diagram gives the
following potential contributions:
\begin{align}
 V^{(q)} = & - \frac{G^3 m_1 m_2^3}{6 r^3} \\
 V^{(s)} = & - \frac{G^3 m_1^2 m_2^2}{r^3},
\end{align}
where the contribution of diagram $(q)$ must be added with
$(1 \leftrightarrow 2)$ in the final potential.

The next set of diagrams at $\mathcal O(G^3v^0)$ are no longer as
simple as $(q)$ and $(s)$. This is due to the presence of the
four-point function, which is used to construct the diagrams $(r)$,
$(t)$, and $(u)$. The only four-point function needed at 2PN is
derived by joining two $\phi\phi\sigma_{ij}$ vertices with an
intermediate $\sigma_{ij}$ propagator, Eq. (\ref{eqn:propsig}), and
attaching $\phi$ propagators, given in Eq. (\ref{eqn:propphi}), to
the external legs. This is given by
\begin{align}\label{eqn:4pt}
 \left<T \underbrace{\phi_a \phi_b}\,\underbrace{\phi_c \phi_d} \right> & =
\frac{i}{128 m_{Pl}^2} \, \delta(t_a - t_d) \delta(t_b - t_d) \delta(t_c - t_d) \notag \\
& \ \ \times \frac{\mathbf k_a \cdot \mathbf k_c \mathbf k_b \cdot
\mathbf k_d + \mathbf k_a \cdot \mathbf k_d \mathbf k_b \cdot
\mathbf k_c - \mathbf k_a \cdot \mathbf k_b \mathbf k_c \cdot
\mathbf k_d}{(\mathbf k_a+\mathbf k_b)^2 \mathbf k_a^2 \mathbf k_b^2
\mathbf k_c^2 \mathbf k_d^2} \, (2 \pi)^3 \, \delta^3 (\mathbf k_a +
\mathbf k_b + \mathbf k_c + \mathbf k_d),
\end{align}
where each set of two $\phi$ fields connected by a brace have the
same intermediate $\phi\phi\sigma_{ij}$ vertex.

Of the three diagrams left to compute, $(t)$ is the most involved,
so we will compute it explicitly. We first note that the symmetry
factor of this diagram is 1/2. The diagram is constructed in the
usual manner, using Eq. (\ref{eqn:worldphi}) and Eq.
(\ref{eqn:4pt}), but we will only work to $\mathcal O(G^3 v^0)$ in
what follows neglecting all velocity factors of the vertices. First,
we integrate over all the time delta functions. Then, we relabel the
three-momentum of the four-point function as $(\mathbf k_a,\mathbf
k_b,\mathbf k_c,\mathbf k_d)\rightarrow(\mathbf k,\mathbf q,\mathbf
p,\mathbf l)$ and couple $\mathbf k$ and $\mathbf p$ to worldline 1
and $\mathbf q$ and $\mathbf l$ to worldline 2. The $\mathbf l$
integration is then performed using the $\delta$-function, which
takes $\mathbf l\rightarrow-(\mathbf k+\mathbf p+\mathbf q)$. After
some algebra, redefining of $\mathbf k \rightarrow \mathbf k+\mathbf
p$ is seen to be useful. This gives the following expression:
\begin{equation}\label{eqn:G3amp_1}
- i \int dt \, V = \frac{i m_1^2 m_2^2}{256 m_{Pl}^6}\int dt \int_{\mathbf k}e^{i \mathbf k
\cdot \mathbf r}\left(I_{\textrm{I}}(\mathbf
k)+I_{\textrm{II}}(\mathbf k)+I_{\textrm{III}}(\mathbf
k)+I_{\textrm{IV}}(\mathbf k)\right),
\end{equation}
where the terms $I_{\textrm{I}}(\mathbf k)$,
$I_{\textrm{II}}(\mathbf k)$, $I_{\textrm{III}}(\mathbf k)$, and
$I_{\textrm{IV}}(\mathbf k)$ are the integrals
\begin{align}
I_{\textrm{I}}(\mathbf k)&=\int_{\mathbf p, \mathbf q}-\mathbf k\cdot\mathbf p \mathbf q^2/Q(\mathbf k, \mathbf p, \mathbf q)\\
I_{\textrm{II}}(\mathbf k)&=\int_{\mathbf p, \mathbf q} \mathbf k\cdot\mathbf q \mathbf p^2/Q(\mathbf k, \mathbf p, \mathbf q)\\
I_{\textrm{III}}(\mathbf k)&=\int_{\mathbf p, \mathbf q}\mathbf p^2 \mathbf q^2/Q(\mathbf k, \mathbf p, \mathbf q)\\
I_{\textrm{IV}}(\mathbf k)&=\int_{\mathbf p, \mathbf q}-\mathbf k^2
\mathbf p\cdot\mathbf q /Q(\mathbf k, \mathbf p, \mathbf q),
\end{align}
and $Q(\mathbf k, \mathbf p, \mathbf q)=(\mathbf k+\mathbf q
)^2(\mathbf k-\mathbf p)^2(\mathbf k +\mathbf q-\mathbf p)^2\mathbf
q^2\mathbf p^2$. These integrals correspond to two-loop Feynman
integrals, and we will show how to compute them efficiently by
reducing them to the one-loop master integral in Eq.
(\ref{eq_FMint1}).

The first three integrals $I_{\textrm{I}}(\mathbf k)$,
$I_{\textrm{II}}(\mathbf k)$, and $I_{\textrm{III}}(\mathbf k)$ can
be computed straightforwardly using the $d$-dimensional one-loop
scalar integral from Eq. (\ref{eq_FMint1}) twice, on both the
$\mathbf p$ and $\mathbf q$ integrations. One can see that
$I_{\textrm{I}}(\mathbf k)=I_{\textrm{II}}(\mathbf k)$ by making the
replacement $(\mathbf p, \mathbf q)\rightarrow(-\mathbf q,- \mathbf
p)$, in $I_{\textrm{II}}(\mathbf k)$. The reason why one can
evaluate these two-loop integrals employing solely a single one-loop
master integral, is that these integrals correspond to nested loop
diagrams. Here, it is important to keep the dimension $d$ of the
integrals arbitrary and only set $d=3$ after the final Fourier
integration over $\mathbf k$ has been performed.

The remaining two-loop integral, $I_{\textrm{IV}}(\mathbf k)$,
requires an integration by parts trick \cite{Smirnov:2006ry} in
order to reduce it in terms of our master integral of Eq.
(\ref{eq_FMint1}). But first, we redefine $\mathbf q \rightarrow
\mathbf q+\mathbf k$. This allows us to split the integral into four
terms, three of which can be evaluated using same methods as
discussed in the previous paragraph. The remaining term is a
two-loop integral with five factors in the denominator
\begin{equation}
\frac{\mathbf k^4}{2}\int_{\mathbf p, \mathbf q}\frac{1}{\mathbf
q^2(\mathbf q-\mathbf k)^2(\mathbf q+\mathbf p)^2(\mathbf p+\mathbf
k)^2\mathbf p^2}.
\end{equation}
Using the integration by parts trick we can write this integral as
\begin{equation}
\int_{\mathbf p, \mathbf q}\frac{1}{\mathbf q^2(\mathbf q-\mathbf
k)^2(\mathbf q+\mathbf p)^2(\mathbf p+\mathbf k)^2\mathbf p^2}=
\frac{2}{d-4} \int_{\mathbf p, \mathbf q}\left(\frac{1}{\mathbf
p^2\mathbf q^2(\mathbf q+\mathbf k)^2(\mathbf p+\mathbf k)^4} -
\frac{1}{\mathbf p^2\mathbf q^2(\mathbf p+\mathbf q)^2(\mathbf
p+\mathbf k)^4}\right).
\end{equation}
Once in this form, we can evaluate it using Eq. (\ref{eq_FMint1})
twice.

After all the Feynman integrals have been evaluated,
we perform the remaining Fourier transformation integral using
Eq. (\ref{eq_FTint}). Subsequently, we can set $d=3$ and find a contribution
to the potential
\begin{equation}\label{eqn:G3amp_4}
V^{(t)}=-\frac{2 G^3m_1^2 m_2^2}{r^3}.
\end{equation}

The remaining two diagrams, $(r)$ and $(u)$, are simpler to compute.
They are evaluated with the methods we have discussed, where once
again it is necessary to compute two-loop integrals by applying the
master integral Eq. (\ref{eq_FMint1}) to two integrations. The
results of the calculations are
\begin{align}
 V^{(r)} = & - \frac{G^3 m_1 m_2^3}{3 r^3} \\
 V^{(u)} = & \ 0,
\end{align}
and the contribution of the potential $V^{(r)}$ must be added with
$(1 \leftrightarrow 2)$ in the final potential. It is worth noting
that $V^{(u)} = 0$ because the diagram gives a purely short distance
contribution $V\sim\delta(r)$, which we have dropped.

\section{Results}\label{sec:Results}

Having computed all the diagrams, we can now construct the 2PN
interaction Lagrangian. There are two contributions which we need.
The first is the kinetic energy, which is obtained by expanding the
matter coupling action, Eq. (\ref{eqn:pp}), to $\mathcal O(v^6)$
while setting all fields to zero. The final potential comes by
summing each potential contribution from the diagrams, $V^{(a)}$ to
$V^{(u)}$. At this stage, we add all contributions with the
worldlines interchanged, as appropriate. Our 2PN interaction
Lagrangian for a binary system is then given by
\begin{align}\label{eqn:2pnlag}
 L_{2PN} = & \frac{m_1\mathbf v_1^6}{16}  \notag \\
           + & \frac{G m_1 m_2}{r} \Bigg(\frac{7}{8} \mathbf v_1^4
             - \frac{5}{4} \mathbf v_1^2 \mathbf v_1 \cdot \mathbf v_2 - \frac{3}{4} \mathbf v_1^2
               \mathbf n \cdot \mathbf v_1 \mathbf n \cdot \mathbf v_2 + \frac{3}{16} \mathbf
               v_1^2 \mathbf v_2^2 + \frac{1}{8} (\mathbf v_1 \cdot \mathbf v_2)^2 \notag \\
           & {} \hspace*{45pt} - \frac{1}{8} \mathbf v_1^2 (\mathbf
              n \cdot \mathbf v_2)^2 + \frac{3}{4} \mathbf n \cdot \mathbf v_1 \mathbf n
              \cdot \mathbf v_2 \mathbf v_1 \cdot \mathbf v_2 + \frac{3}{16} (\mathbf n \cdot
              \mathbf v_1)^2 (\mathbf n \cdot \mathbf v_2)^2 \Bigg)\notag \\
         + & G m_1 m_2 \Bigg(\frac{1}{8} \mathbf a_1 \cdot \mathbf n
           \mathbf v_2^2 +  \frac{3}{2} \mathbf a_1 \cdot \mathbf v_1 \mathbf n \cdot \mathbf
           v_2 - \frac{7}{4} \mathbf a_1 \cdot \mathbf v_2 \mathbf n \cdot \mathbf v_2 -
           \frac{1}{8} \mathbf a_1 \cdot \mathbf n (\mathbf n \cdot \mathbf v_2)^2 \Bigg) \notag \\
         + & G m_1 m_2 r \Bigg(\frac{15}{16} \mathbf a_1 \cdot
            \mathbf a_2 - \frac{1}{16} \mathbf a_1 \cdot \mathbf n \mathbf a_2 \cdot \mathbf
              n \Bigg) \notag \\
         + & \frac{G^2 m_1 m_2^2}{r^2} \left(\frac{7}{4} \mathbf
             v_1^2 + 2 \mathbf v_2^2 - \frac{7}{2} \mathbf v_1 \cdot \mathbf v_2 + \frac{1}{2}
             (\mathbf n \cdot \mathbf v_1)^2\right) \notag \\
         + & \frac{G^3 m_1 m_2^3}{2 r^3} + \frac{3 G^3 m_1^2
             m_2^2}{2 r^3} + (1 \leftrightarrow 2),
\end{align}
where $(1 \leftrightarrow 2)$ refers to all terms given previously, with
the labels 1 and 2 interchanged. Note that $\mathbf n
\rightarrow-\mathbf n$ under this exchange.

Upon comparison to the 2PN piece of the Lagrangian in Eq. (174) in
\cite{Blanchet:2002av}, it is apparent that our Lagrangian does not
have the same form. However, when the equations of motion (EOM) are
computed we recover the same 2PN EOM as in \cite{Blanchet:2002av}.
This shows the physical equivalence of our Lagrangian and the one in
\cite{Blanchet:2002av}. At the level of the Lagrangian, this can be
shown directly by relating our result in harmonic gauge to the
standard 2PN Lagrangian in harmonic gauge of \cite{Blanchet:2002av}
via a total derivative and two double zero terms
\cite{Barker:1980dblz}. A double zero term is an expression which
vanishes at 2PN by the use of the lower order EOM. Adding such a
term modifies the form of the Lagrangian, but does not change the
EOM or the choice of gauge. The terms we add are given by
\begin{align}\label{eqn:dzF}
 \delta L_1 = & \frac{1}{8}\frac{G m_1 m_2}{r}\left(\mathbf r\cdot \mathbf
a_1+\frac{G m_2}{
r}\right)\left(\mathbf r\cdot \mathbf a_2-\frac{G m_1}{r}\right)\\
\delta L_2 = & - \frac{15}{8}G m_1 m_2 r\left(\mathbf
a_1^i+\frac{G m_2}{r^3}\mathbf r^i\right)\left(\mathbf a_2^i - \frac{G m_1}{r^3}\mathbf
r^i\right)\\
\delta L_3 = & \frac{d}{dt}\left[\frac{7}{4}\frac{G^2 m_1 m_2}{
r^2}\left(m_2 \mathbf r\cdot \mathbf v_2-m_1 \mathbf r\cdot \mathbf
v_1\right)+\frac{3}{4}\frac{G m_1 m_2}{ r}\left(\mathbf r\cdot
\mathbf v_1 \mathbf v_2^2 - \mathbf r\cdot \mathbf v_2 \mathbf
v_1^2\right)\right],
\end{align}
where the first two terms are double zero terms and the last one is
a total time derivative. By adding these terms to our Lagrangian at
2PN, Eq. (\ref{eqn:2pnlag}), we obtain the Lagrangian in
\cite{Blanchet:2002av}. We again emphasize that applying these
transformations leaves the EOM unchanged, so that we remain in
harmonic gauge.

We have also calculated the observable $E(\omega)$ for a circular
orbit and recover the known 2PN result \cite{Blanchet:2002av}. We
note that if the accelerations were replaced in our Lagrangian,
using the LO EOM, $E(\omega)$ is unchanged, although all
intermediate expressions are different. This is because $E(\omega)$
is physical and therefore gauge invariant, and using the EOM at the
level of the Lagrangian corresponds to a change in gauge.

\section{Conclusion}\label{sec:Conclusions}

We have demonstrated how to use the EFT method
\cite{Goldberger:2004jt} to efficiently calculate the
next-to-next-to-leading order Lagrangian describing the conservative
dynamics of a binary system. We have shown how to systematically
determine all Feynman diagrams which contribute to the 2PN
Lagrangian through the use of power counting in $G$ and $v^2$. The
calculation involved 21 distinct diagrams, and we encountered
integrals which correspond to one-loop and two-loop Feynman
integrals. All diagrams could be computed with only two master
integrals. This demonstrates the efficiency of the EFT method for
the calculation of the 2PN Lagrangian.

Instead of a usual Lorentz covariant metric parametrization, we
employed a temporal Kaluza-Klein parametrization of the metric,
proposed by Kol and Smolkin \cite{Kol:2007rx, Kol:2007bc}. When
compared to the $h_{\mu\nu}$ parametrization used in
\cite{Goldberger:2004jt}, we found that the KS variables reduced the
number of diagrams by four at $\mathcal O(G^3v^0)$ and avoided any
propagator insertions in the three-graviton diagrams at $\mathcal
O(G^2v^2)$. This reduced the amount of calculation significantly at
2PN. More generally, we conclude that the KS variables will
significantly improve any calculations which are performed at higher
order using NRGR methods. This is due to (1) the simple propagators
obtained with this gauge and parametrization choice, (2) the
suppression of $\phi^n$ vertices by one order in the PN expansion
due to two time derivatives acting on the vertex, (3) not requiring
time insertions in the $\phi^n$ topologies at the next order in the
PN expansion, and (4) the advantageous compounding effect of the
former points when higher order expansions are calculated.

An interesting extension of the current calculation is to compute
the 3PN potential. Analogously to 2PN, where the four-point function
formed diagrams which were the most difficult to evaluate, the
hardest part of the calculation at 3PN is expected to involve a
gravitational four-point function in $\mathcal O(G^4v^0)$ diagrams.
Also, the number of diagrams one must evaluate at 3PN is much
greater. This means more terms in the action will have to be
computed to account for new vertices at $\mathcal O(G^4v^0)$,
$\mathcal O(G^3v^2)$, $\mathcal O(G^2v^4)$, and $\mathcal O(Gv^6)$.
Our preliminary estimates show there are more than $100$ diagrams at
3PN, compared to the $21$ at 2PN. At 3PN, it will again be the case
that a reduction in the number of diagrams can be achieved if the KS
variables are employed. So while the 3PN calculation may prove
challenging, the relative ease of the calculation at 2PN indicates
that the NRGR calculation of the 3PN potential can be done.

\begin{center}
{\bf Acknowledgements}
\end{center}

The authors thank Walter Goldberger for many insightful discussions
and comments. This work has been supported in part by Grant No.
DE-FG-02-92ER40704 from the US Department of Energy.

\end{document}